\documentstyle[12pt,aasms4]{article}
\begin{document}
\title{JHK imaging and photometry of low z QSOs and radio galaxy}

\author{J.B.Hutchings\altaffilmark{1}}
\affil{Dominion Astrophysical Observatory, National research Council of
Canada, 5071 W. Saanich Rd., Victoria B.C. V8X 4M6, Canada}
\author{S.G.Neff\altaffilmark{1}}
\affil{Laboratory for Astrophysics, Code 600, NASA Goddard Space Flight
Center, Greenbelt, MD20771}
\authoremail{hutchings@dao.nrc.ca,neff@stars.gsfc.nasa.gov}
\altaffiltext{1}{Guest Observer, Canada France Hawaii Telescope,
which is operated by NRC of Canada, CNRS of France, and the University 
of Hawaii}
\begin{abstract}
 We describe J,H,K deep imaging of 90 arcmin fields around 4 QSOs and one
Radio galaxy at redshifts in the range 0.06 to 0.30, and show their images,
luminosity profiles, and NIR 2-colour diagrams of objects. We find
that the QSO hosts are all resolved, and compared them with previous CCD images. 
The host galaxy colours are consistent with old and young stellar populations 
at the QSO redshift. The colours of nearby galaxies suggest that all the AGN
live in groups of generally smaller companion galaxies, mostly with evolved
populations at the same redshift. The two radio-loud objects live in richer
cluster environments than the others. Gissel population models indicate
reddening in the galaxies, star-forming regions, and possibly a systematic 
H-K offset. The QSO luminosity profiles are complex and reveal some of their
tidal disturbance and star-formation history. 
\end{abstract}

\section{Introduction}

   The host galaxies of QSOs, and radio galaxies at low redshift 
(approximately 0.1 to 0.5), have been resolved and studied quite extensively
at optical wavelengths (e.g. Hutchings Jansen 1987, Hutchings and Neff 1991,
Bahcall et al 1995). At these wavelengths the images are dominated by the
nuclear light and also by young stellar populations. In many cases the host
galaxies are interacting systems which appear irregular because of tidal
distortions, as well as associated star- and dust-formation. The CCD images 
have revealed much about these processes and the emission-line gas associated
with the galaxies.

   In recent years it has become possible to obtain NIR images of faint objects
with good resolution and efficiency. Compared with optical wavelengths, 
the NIR offers the advantage of a lower 
brightness contrast between the active nucleus and the host galaxy stellar
population, and as redshift increases it offers a lower 
k-correction of the host galaxy. Dunlop et al (1993) and McLeod and Rieke
(1994) have published NIR imaging investigations of QSOs. 
This paper describes extension of this work in the NIR in several ways.

   First, we have obtained images in 3 colours: J, H, and K, enabling the
derivation of colour-colour plots which are diagnostic of stellar populations.
We have compared the data with Gissel models (Bruzual and Charlot 1993),
which are sensitive to both old and young stellar populations. Second, 
we have chosen a small sample of objects for very deep imaging, all of 
which have previously been resolved well at shorter wavelengths: thus 
they are selected for known host galaxy properties. Third, we have used 
a wide field so that we can study the properties of nearby (possible
companion) galaxies, to study their group properties and stellar populations.

\section{Observations and data analysis}

   The observations were obtained with the Redeye wide-field camera at the 
prime focus
of the Canada France Hawaii telescope in January 1996. The field size covered by
the HgCdTe array is 2 arcmin, sampled with 0.5 arcsec pixels. This camera
increases field size and pixel signal level at the expense of spatial
resolution. The image quality during the run was estimated at 0.8 to 1.0 
arcsec, so that the
images are somewhat (but not badly) undersampled. Total exposure times with each
filter were approximately 32 minutes, which gives us 3 colour photometry with
good accuracy to magnitudes ranging from 19.5 in K to 21 in J 
(see discussion below). Detection limits are some magnitudes fainter.

   Observations were made using dither rasters with 10" (occasionally 15")
offsets in 4 x 4  or 3 x 3 square patterns. These allow us to obtain good 
sky and flat field frames without taking extra observing time, with a fully
exposed combined field of 90 arcsec. The observations were obtained on two
nights: conditions were clear and calibration images of standard stars were 
obtained on each night. The K' filter was used and calibrated with K
magnitudes, so that we simply quote K magnitudes in our results.
Individual frame integration times
ranged from 20 seconds to 120 seconds in J and 15 seconds
to 60 seconds in K, to keep sky signal levels safely below detector saturation.
For the shorter exposures, several exposures were taken at each dither 
position before moving on. Dark frames were obtained but at these exposure
levels were not significant. Dome flats were also recorded. 
Table 1 summarizes the observations and lists the targets.

   The data were processed as follows. Sky frames were obtained
by median combination of successive dither images. The exact dither
configuration and sequence, and the rate of change of sky brightness, 
determined the optimum number to combine: typically the sky was stable 
enough to combine 20 minutes worth of data. In these intervals the
raw exposures were scaled to a constant sky level (typically by $<$5\%,
and no more than 12\%) and the median sky subtracted from them. With the short
individual exposures we used, dark frames simply added noise and were not
used. Flat fielding also produced no measurable difference for these
data, again being a source of noise. In the intervals of more rapid or
extensive sky brightness changes, the flats and more frequent sky frames were
necessary. Figure~\ref{mosaic} shows subimages centred on the QSOs and Radio Galaxy, 
showing that they are all clearly
resolved. We discuss them in detail in the sections below.

   The resulting images have very similar quality and S/N, since all
received essentially the same exposure. We estimated the noise and signal
level in several ways: from the blank regions in the final sky-subtracted
images; from the measured signal and noise levels in the individual frames;
from the instrument performance values of read noise and sensitivity; and
from individual frame measures of bright target objects and standard stars.
These estimates agreed to well within their values.
Overall, we achieve 1$\sigma$ errors in magnitudes of 0.1 at 18.8; 19.6;
20.3 in K, H, and J. At 0.2 mag error, the magnitudes are 19.9, 20.7, 21.5
respectively. Faint limits vary somewhat between objects depending on their
location in the detector (there were reads at four corners) and some flat 
field features. Thus, in the discussions that follow, we include only 
brighter objects with magnitudes good to 0.15 in all three filters. 
Generally this is some 20 objects per field. In the faintest images 
(usually J-band) we can detect at least 2 times this number of objects. 
The fluxes were measured using the IRAF\footnote{IRAF is distributed by 
NOAO, which is operated by AURA Inc., under cooperative agreement with 
the NSF.} task Imexam using different object 
and background radii as required. In all cases the radii used were those
where the signal is less than the noise in the azimuthally averaged profile. 
In the few cases of crowding of
images, the blended images were removed by editing, or the object(s) rejected.

   In the sections below we describe the detailed results for all 5 of the objects. For all of them we have performed the following measurements. 
First, we have
measured J, H, and K magnitudes for all objects in the fully exposed fields
with measuring error 0.15 in all bands, and plotted H-K/J-H diagrams for them,
with error bars. We have removed all identified stars from these plots: stars
are generally easily identified in a plot of peak signal against total, as
the sampling places most of their flux in a single pixel. To the limits of
our interest, there are typically 5 stars per field. One low latitude field
has more stars. We have also measured all or parts of the QSO host galaxy,
separated from the nuclear light. In all cases this was possible, and in some
we were able to measure different parts or features of the host galaxy. 
Finally, we derived the azimuthally averaged or ellipse-fitted profile 
of the QSO or RG using the IRAF task Ellipse. Table 2 shows all the measures
used in this paper. 

\section{Stellar population models}

Model values for the J, H, K fluxes were calculated using the
Gissel stellar population models (Bruzual and Charlot 1993), and 
compared with the measures for each field and appropriate redshift.
Figure~\ref{mod} shows some of these
tracks on their own, and later figures compare specific ones with the data.
The models are all for a passively evolving population after a 1 Gyr starburst.
The Padova initial spectral energy distribution (SED) is used
with a Salpeter initial mass function (IMF). Other models produce essentially
the same results. Continued star-formation, renewed star-formation, or
a truncated IMF all tend to move galaxies nearer the start of the tracks.

The H-K colours are sensitive to whether the K or K' bandpass are used, and
the models are adjusted to the CFHT K' filter. However, mismatches in
bandpasses may cause the model colours to have systematic offsets of up to
0.1 in J-H and H-K. 
The tracks evolve very quickly from their lower left to the upper portion 
after cessation of star-formation. The top end of the tracks occupies most 
of the time, as seen from the time marks: thus an evolved population
will fill a small area of the diagram that moves with redshift, faster with 
increasing redshift. At the redshifts of interest in this paper (0.06 to 0.30)
we are able to distinguish objects with a resolution of about 0.03 in z 
for bright objects, and twice that at the faint limits of the measures we
have used. Since any field will contain background and foreground
objects, we can use the diagram to determine which are at the QSO redshift, if
they are old populations. 

If star-formation is extended or renewed, objects will lie nearer the lower 
left of the tracks: thus we can identify objects with younger populations,
since no other circumstances can move them lower in the diagram. Unfortunately this 
is not true of all objects in the upper parts of the diagram, since reddening
moves them along almost the same locus as redshift increases above z = 0.15.
Thus, we need to consider evidence for dust in interpreting the data.
 
Some cautionary remarks are necessary on the models. They may be poorly defined
in the NIR, so that there may be systematic errors in them, particularly at lower
redshifts. Thus, we may use them as guidelines but be wary of conclusions
that seem unusual, such as the reddening levels suggested in the sections
below. However, irrespective of the models, the positioning of objects in the
diagram, particularly grouping of objects, is good evidence for having similar
stellar populations and redshifts.

\section{0052+252}

   This z=0.16 radio-quiet QSO has been imaged at optical wavelengths by
Hutchings and
Neff (1989) using the uncorrected CFHT, and by Bahcall, Kirhakos and Schneider 
(1996: BKS) using the corrected Hubble Space Telescope. 
The QSO was discovered optically
by the Schmidt and Green survey (1983), and has since been observed to have a
weak radio flux (Kellerman et al 1995). Dunlop et al (1993) 
obtained K-band images and MacLeod and Rieke (1994) obtained 
H-band images. The summary of previous observations (of which the HST images
naturally have the most detail), is that the host galaxy looks like an inclined 
spiral with a pronounced outer ring of what look like irregular H II regions.
This is supported by [O III] imaging by Stockton and MacKenty (1987) which
indicate the presence of ionised gas in the outer disk.
There are two faint stars within the projected host galaxy which confused the
lower resolution images. The QSO has several bright galaxy companions of lower
luminosity and mostly of elliptical appearance. 

   Figure~\ref{f00im} shows the J band image of the field, with measured objects identified.
The brighter galaxies are labelled A to G following the nomenclature of
BKS. In Figure~\ref{f00pl} we show the 2-colour NIR diagram of these objects. The QSO
host galaxy colours were measured in several ways: individual regions which
are well separated from the nucleus were measured and corrected for local sky
level; the difference between nuclear and whole galaxy measures was used; and
the differences between azimuthally averaged plots of the QSO at large radii.
The results are consistent with the point marked and its error bars shown.

  Figure~\ref{f00pl} also shows the model tracks for a stellar population as described
above, for some redshifts. The z=0.16 redshift model is also shown with an
amount of reddening that appears to fit the data for the QSO and bright
companions (i.e. the track that has 2-10 Gyr old stars at this position in 
the diagram). The bright galaxies A, B, C, D all lie in a close group in the
diagram, together with the QSO host galaxy. The fainter galaxies E, and F 
lie further away on the diagram but their measuring error bars are consistent 
with their being associated too. G lies further away in the sky, is clearly 
a barred galaxy in the HST image, with somewhat larger scale length: this
morphology and its position in the diagram are also consistent with it being 
a lower redshift (foreground) galaxy. A few other galaxies lie in this area 
of the diagram: 1, 7, 15, 16, 18, and 19. These lie further from the QSO 
and are fainter than the bright ones. The results suggest that the bright
galaxies form a compact group associated with the QSO. The QSO host is the
largest and the only one with structure: the others have the size and 
luminosity of small ellipticals, somewhat like the M31 group. 

   With the exception of G and 17, no galaxies suggest a young population. Most
of the faint galaxies have the colours of redshift 0.4 and higher. The QSO
nucleus has colours that lie to the right of any model tracks, reflecting 
a non-stellar SED.

   The overall group colours are not consistent with unreddened models. They 
lie close to the positions predicted for A$_v$ values of 2-3 magnitudes. We 
do not have optical colours to check whether this reddening is indicated in 
the visible. If it is real, there must be dust spread through the group, since
it is so similar for all the galaxies. In this, and the other cases
below, it seems probable that the reddening indicated is not real,
but an indication of systematic errors in the IR models.

   The QSO luminosity profiles are shown in Figure~\ref{f00pr}. 
We have included the
(approximately R-band) image from the HST archive, smoothed to the sampling
of our NIR images. These profiles were derived from edited subimages with 
the two stars and nearby companions removed. The profiles show that the 
NIR flux falls faster than the V-band, indicating an old population
that is more centrally condensed than the young population seen in the ring 
of H II regions and the extended [O III] emission. We find that the profile 
is not well fit by either an exponential or r$^{1/4}$ law. The NIR images 
show no signs of the bright H II regions, and the host galaxy projected
major axis is tilted
somewhat from the visible as a result. There is no evidence of tidal tails 
of old stars, although the NIR images are not as deep as the CFHT R-band, 
which show the QSO host extends as far as galaxies A and B. The NIR images 
do show some faint extended signal about 70\% of the distance to both 
galaxies (more evident in J and K than H) which coincide roughly with the 
[O III] seen by Stockton and MacKenty (1987). Thus there may be some emission
(e.g. Paschen lines) at NIR wavelengths too. 

   Overall, there is no sign of any very recent tidal events, but the QSO 
host is clearly in a stage of secondary star-formation. The presence of 
an associated compact group of galaxies and the luminosity profile
suggests that a tidal event may have triggered the QSO and star-formation.

\section{0157+001 = Mkn1014}

    This is a radio-quiet QSO in the Schmidt and Green (1983) sample, at
redshift 0.16. Stockton and MacKenty (1987) published [O III] images, and
MacKenty and Stockton (1984) 
and Hutchings and Crampton (1990) discussed images and off-nuclear spectra.
The QSO resides in a galaxy with strong spiral features in a near face-on
aspect. The arms have blue colour and line emission is seen (which may not
follow the arms closely), so that there
appear to be regions of recent active star-formation.

    Figure~\ref{f01im} shows one of our images with numbered companions that we have 
measured. There are many faint small objects near the QSO: in particular
objects 5 and 17 appear at the end of the two major spiral features, and 3 lies
in a dense region of one arm. It seems very probable that these are condensations
within (tidally drawn?) major arms or tails of the host galaxy. They are more
prominent in the NIR than at optical wavelengths.

   Figure~\ref{f01pl} shows the 2 colour diagram of the objects in the field, together
with some relevant model tracks. We note that no objects fit the unreddened
models for z=0.16 (or any other redshift). The best fits indicate the model shown,
which has A$_v$=3 at the rest wavelength, and a local (z=0) A$_V$=0.3. 
 We have measured the QSO nucleus and whole (nucleus plus host galaxy),
and these are marked. We also measured the colours of the two major arms (A, the 
lower one - excluding object 5, and C, the upper one, and the brightest region within A, as B).
The objects fall in two main groups in the diagram. The arms and associated object
17 appear to belong to a younger population, along with objects 1, 4, and 7, all of
which are close to the QSO. With the exception of object 8, all others appear
to have an older population. As for 0052+253, the QSO nucleus lies to the right
of the group in the 2-colour plot, while the total flux (nucleus plus host) 
has old population colours. Within the host
galaxy, the object 5 has high J-H colours while the others, 3, 4, 17 are low
enough to be young population candidates. 

   The luminosity profiles (Fig~\ref{f01pr}) show the mixed properties that are common in QSOs
(see Hutchings and Neff 1991). The diagram includes the PSF from the 
bright nearby star, as this is our best illustration of a typical PSF.
The QSO profile is well resolved outside of $\sim$1.8 arcsec.
There is a good fit to an r$^{1/4}$ profile
in the inner 5 arcsec, which is the bright inner galaxy. From radius 5 to 17
arcsec there is a reasonable exponential profile, as suggested by the 
spiral structure seen
in this range. The exponential slope is steeper with increasing wavelength,
while the inner bulge slope is the same in all our bands. 
 The bright features at the end of the arms suggest that these
are tidal features rather than features of a normal disk galaxy, and the high
concentration of faint galaxies of old population that appear to be real and
close QSO companions suggest that there has been a merger or close encounter.

\section{0453+22 = 3C132}

   This object is a strong double-lobed radio source identified with a galaxy
at z = 0.21. Optical range imaging was reported by Hutchings, Jansen, and Neff
(1987), and the images are shown in Hutchings, Johnson, and Pyke (1988). The
galaxy is in a crowded field and its faint outer parts extend over a number of
superposed faint objects. The object is at fairly low galactic latitude, so
the field contains more stars than the others in the program. Those
identified unambiguously from peak/total flux ratios were ignored and there
were no dubious cases. Figure~\ref{f04im} shows the field and measured objects. 

    The field is more crowded than those of the QSOs, and almost all of the 
measured galaxies form a tight group in the 2 colour diagram around the radio
galaxy (Figure~\ref{f04pl}). 
Thus, we have evidence for a group or cluster around the radio source, which
is richer than those around the QSOs. Unfortunately the two brightest nearby
galaxies, which are seen but not marked in Figure~\ref{f04im}, could not be measured
in our K-band image, because of a small telescope pointing offset with respect to the other 
observations. However, they may be the more luminous members of this cluster.
The radio galaxy has no distinct nucleus at our resolution, but
was measured for colour
in two spearate regions away from the centre and the colours are within
measuring errors the same as the whole galaxy. Thus, the radio galaxy does
not contain a nuclear component with different SED from the whole, again unlike the
QSOs. 

    Figure~\ref{f04pl} shows the loci of 10 Gyr age galaxies at the radio galaxy redshift
for two values of reddening. The galaxies in this field are more `reddened' than
the QSO fields.

   The galaxy profile is difficult to measure because of the overlapping stars
and companions. However, the profiles obtained from edited images in which the
compansions are removed are all very similar. This profile does not fit either
and exponential or de Vaucouleurs profile in any radius range out to 10 arcsec
where our signal vanishes. Thus, again the galaxy appears to be unusual, and possibly
in a recent post-interaction state. The close-by objects 11 and 13 could be
interacting companions, or the galaxy could have destroyed a smaller companion
in a recent merger. 

\section{0844+439}

   This is a low redshift (z=0.064) QSO from the Schmidt and Green survey (1983). Its
luminosity and easily detected host galaxy have caused it to be excluded from
the Hewitt and Burbidge (1993) catalogue. An image and spatially resolved
spectra were published by Hutchings and  Crampton (1990). The host galaxy is
a barred spiral with an unusual bright outer ring, and it lies close to
a large and bright companion which looks like an inclined spiral. There are
several fainter non-stellar objects close to the main pair of galaxies.
The disturbed QSO host is clearly in a recent post-interaction state. The off-nuclear
spectra show the presence of ionised gas and a young stellar population.

   Figure~\ref{f08im} shows the field and the measured objects. This is the most sparsely
populated field in the sample. The 2 colour plot (Figure~\ref{f08pl}) contains models
for the QSO redshift (0.064) and the unreddened and reddened loci for an old galaxy.
As for the other QSOs, the QSO lies to the right of the model but the host
galaxy free of the nucleus lies near the expected position with some reddening
(A$_V\sim$1.5?).
The bright companion appears to have the same population as the host galaxy. 
As these are disk
galaxies they may correspond to models of younger population than other objects
in the field. The distribution of objects suggests that many of the objects are
background; some may be in a group at z$\sim$0.2-0.3. Apart from its bright
companion the QSO may be associated only with a few companions (numbers 7, 9, 11,
12 are candidates). Object 10 lies at the end of the tail or arm of the companion
galaxy, and thus may be a new dwarf galaxy forming, as proposed for such
objects e.g. by Mirabel, Dottori, and Lutz (1992).
While it has large error bars, possibly strong line emission could contribute to
its odd colours. 

   The visible image of Hutchings and Crampton contains a bright compact object
in the middle of the outer shell that lies between the QSO and companion which is
not detected in any of our NIR images. Unfortunately their spectra did not
include the object. The strong change between the R and J bands
implies a very strange SED, or that the object has changed. Thus, it is possible that
the object seen by Hutchings and Crampton was a supernova: its measured R magnitude
corresponds to M$_V$ of -17.6 without correction, for H$_0$=100. If the extinction
estimate of 1.5 is right, then it has M$_V$ of $\sim$-19. These are consistent
with supernova luminosities. 

   The luminosity profiles are shown in Figure~\ref{f08pr}, 
from images with the companions 
edited out. As anticipated from the irregular morphology, the profile is neither
exponential nor power law with radius. The regions from radii 1.4 to $\sim$4 arcsec 
are roughly power law, while from 2.5 to 7 or more arcsec an exponential is
a rough fit. The divergence between the colours at large radii is a reflection
of sky frame uncertainties, since the large size of the two main galaxies
gave rise to some `holes' in the median sky frames from the 15 arcsec dither
pattern of the observations. 

  Overall, this QSO shows a high degree of interaction, with one bright neighbour
galaxy, and an otherwise poor environmemt. There is little reddening in the two
main galaxies, and little sign of any additional old population not seen in the
R band image. 

\section{0911+053}

   This is a radio loud QSO, at z=0.3 the highest redshift object in the present
sample. Previous optical imaging is described by Hutchings et al (1987,88). The
QSO has a bright host galaxy which is very close to a companion of similar size
and brightness. Both galaxies have off-centred nuclei, suggesting that there is
a tidal interaction. A fainter elongated galaxy is extended along the QSO galaxy
axis and the inner host galaxy isophotes are extended in its direction. Thus it
may possibly be a tidal remnant.

    The most remarkable feature of the NIR images is that the inner host galaxy
is not just elongated - it is clearly resolved into two separate `nuclei' about
2 arcsec apart, with a flux ratio in the range 5 to 10 (see Figure~\ref{mosaic}). 
The difference arises
from the change in relative optical and NIR flux between the two objects, since
the CCD data have the same image quality. We have attempted to measure the
second nucleus separately in our images. 
 
Figure~\ref{f09im} shows the field and measured objects. As usual, we have not included
fainter objects or those which do not have measures in all 3 colours. Unfortunately an observing offset excluded the next brightest nearby galaxy
(at the bottom centre in Figure~\ref{f09im}) from the H band image. We
have measured the colours of the QSO host galaxy well away from the nucleus,
and also the companion extended flux. Figure~\ref{f09pl} shows the 2 colour plot, in which
the QSO host is designated Hs, and the companion extended light as 12F. The
QSO secondary nucleus is 14.

   In this field the 2 colour diagram has a well-populated group of objects 
near the QSO and companion. Outliers of the group are galaxies 11, 15, 16. 
Other galaxies that appear to be higher redshift
background objects are all further from the QSO in the sky. The QSO group is
well-populated and again suggests that radio-loud objects live in a richer
galaxy environment. The positions of the group in the 2 colour diagram
corresponds with a reddening of A$_V\sim$1 at the QSO redshift, and the
closeness of the group suggests that there are not strongly star-forming
populations. (The lower `reddening' in this, the highest redshift object,
also suggests that it is a measure of model systematics, which are better
in the shorter wavelength regions.)
 The offset of the QSO host makes this the strongest candidate
for containing a younger population of stars. The position of the second
nucleus suggests that it may be more reddened but the offset is small. 
The possibly distorted galaxy 16 (see below) is also a weak candidate
for a young star population.
 
    The luminosity profile of the QSO is not easy to measure since it has
a secondary nucleus. However, the data points away from this follow an
r$^{1/4}$ law which matches that of the companion quite well. Since both
galaxies are extended asymmetrically this is not a very quantitative
measure. However, neither galaxy has any part with an exponential profile,
and it seems probable that they are ellipticals that have had some mild
tidal distortions. The QSO second nucleus is likely a merging object since
it is brighter than other faint galaxies nearby and there are few stars in the
field at all. It also lies along the line of extension of the galaxy
and near the line to object 16. The visible images suggest that 16 may lie
at the end of a tidal tail from the QSO, or is also tidally extended by
its QSO companion. The NIR images do not show any old population of stars in 
this direction, so we favour the latter possibility. 

\section{Discussion}

   There are several principal results, which we list and then discuss. 1) The
QSO hosts are all resolved and can be compared with the CCD images. 2) The
host galaxy colours are consistent with stellar populations - both old and
young - at the QSO
redshift. 3) The colours of nearby field galaxies suggest that all live in
groups of (mainly smaller) companion galaxies with (mostly) evolved
populations at the same redshift. 4) The models indicate reddening in
the galaxies, and also probably a systematic offset. 5) The QSO luminosity
profiles are non-standard and reveal some of their tidal disturbance and
star-formation history. 6) The radio-loud objects live in richer cluster
environments.

   All of our 2-colour plots show grouping of objects with the QSO, which
lie close to the expected position for the QSO redshift in the model tracks.
However, in all cases there is an offset from the unreddened models. We
have discussed these in terms of reddening, but in several cases this implies
a large amount of extinction that may not be likely to apply to the entire
group or cluster. Thus, as discussed in section 3, it is possible that
there is a systematic offset between the models and the measurements,
due to mismatch of the passbands in each. Models with different
initial SEDs are not different by this amount. It is thus possible 
that the reddening values need to be
decreased by a factor of 2-3 in A$_V$, increasing with redshift, 
since a bandpass effect is non-linear in z.
While the absolute values of reddening are uncertain, the relative effects
within each field are determined only by the measuring errors plotted.

 The objects in this sample were chosen as having host
galaxies easily detected in the optical (CCD) range, since we wanted to study their morphology and
stellar populations. The deep NIR images in reasonable seeing were successful
in resolving the hosts in all 3 bands. We have not attempted to remove
the PSFs as a) the images are undersampled, b) there are no good PSF stars
in the data frames, and c) the host galaxy signals are good enough to measure
well away from the nucleus. The data are not well enough sampled (or resolved)
to warrant major PSF removal work, or morphology studies near the nucleus.
In the cases of 0050+252 and 0157+001 the host galaxy structure is blue and
the NIR images show that there are not other old population features present.
The luminosity profiles change with wavelength showing the radial structure
of the stellar population: the central regions are redder/older. We do not
see any new structures that were not visible at CCD wavelengths. The sky
brightness is the limiting factor rather than the actual SED: a darker
NIR sky would perhaps allow us to detect new morphological features.

 The NIR colour of the QSOs, their host galaxies, and companions are good 
diagnostics of their redshift, stellar composition, and possibly dust content.
We have shown how these are traced with Gissel population models, and that
they generally fit well with groups of associated galaxies at the QSO redshift.
The diagrams also reveal background galaxies which fit higher redshift models.
Although for clarity of the main display we did not plot the results for
stars, they lie along the tracks for zero redshift. While reddening and
redshift move the tracks along parallel paths in some redshift regimes, all
comparisons with models indicate a significant reddening, and also an offset
in H-K for the known QSO redshift. Since the reddening implied is larger than
seems reasonable (several magnitudes in V-band), we suggest that there
are systematic effects in the colour indices calcluated by the Gissel models,
of order 0.1 magnitudes. However, the observational clumping of objects
with the QSO or its host galaxy, strongly indicate that there are groups of
associated galaxies in all fields. Few of the objects lie at earlier ages
in the model tracks, compared with the QSO, so that the stellar populations
have probably evolved passively since the group formation. There are a few
exceptions, usually within the host galaxies, where there appears to be
current new star-formation. In some of these this is also seen in
spectroscopic evidence. 

 We have estimated the richness of the companion field roughly by counting
numbers of objects down to our brightness limits and seeing how many of these
appear to be in the QSO group. While this needs to be compensated for redshift
differences, limiting magnitudes, and unmeasured objects due to crowding
and image flaws, the qualitative result is that the two radio-loud objects live
in richer galaxy environments than the others, as has been noted by other
investigations (e.g. Yee and Ellingson 1993). 

  The luminosity profiles probe the older population better than those from
CCD wavelengths. However, none of the objects has a standard exponential
or power law profile, stressing that the whole structure of the galaxy has
been disturbed by the tidal events that are visible to different degrees.
Thus, peculiar morphology does not arise only from new star-formation or
the nuclear radiation in a galaxy: it is seen in the older stellar population as
well. The host galaxies are more centrally condensed with increasing
wavelength in the J, H, K passbands.

\clearpage
\centerline{Captions to figures}

\figcaption[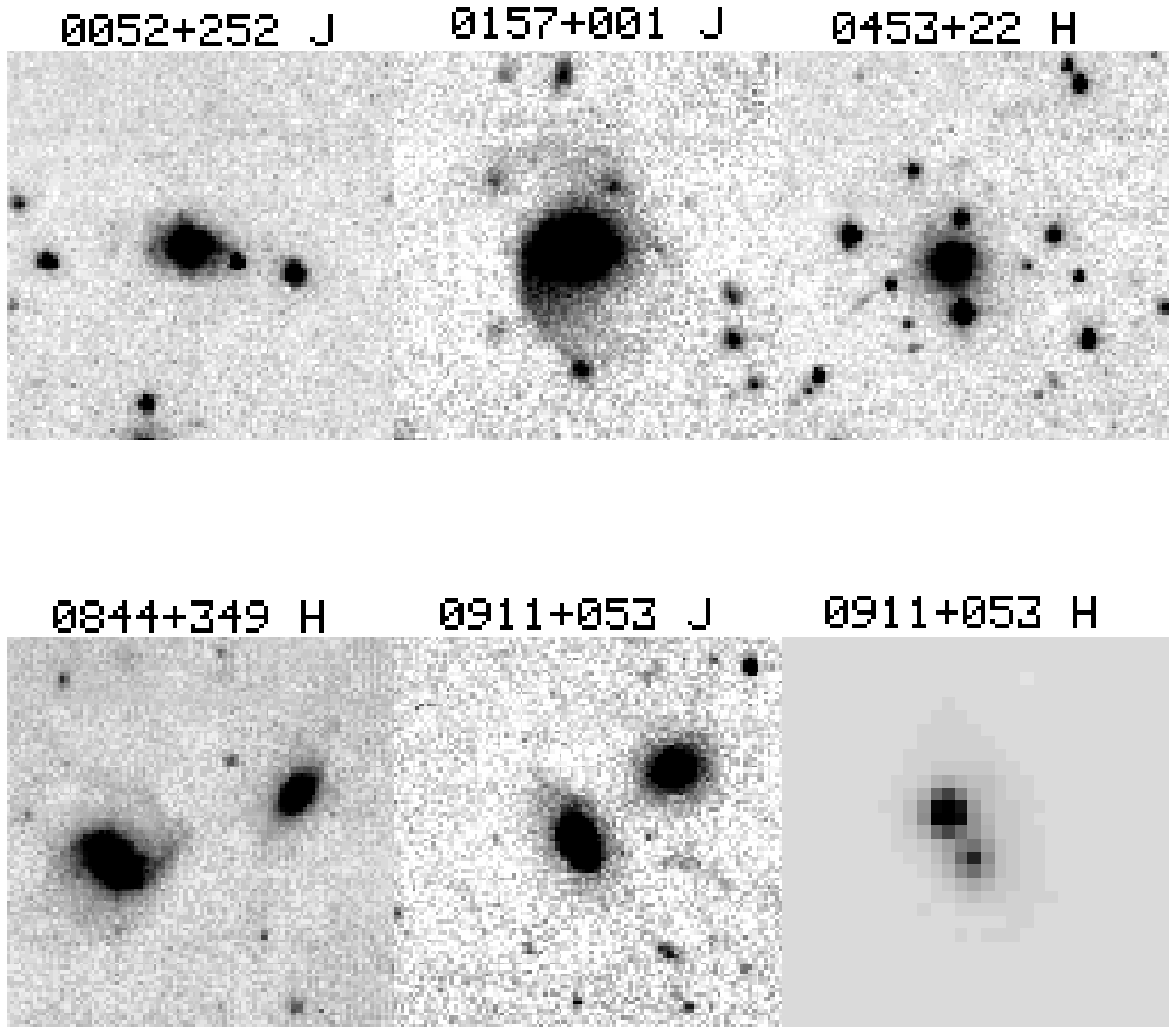]{50 arcsec subimages around the QSO or radio
galaxy from the fields observed. The QSO 0844+349 is the object to the
left in that image. The 0911+053 image is shown with two different
stretches and scales to illustrate the double `nucleus'. \label{mosaic}}

\figcaption[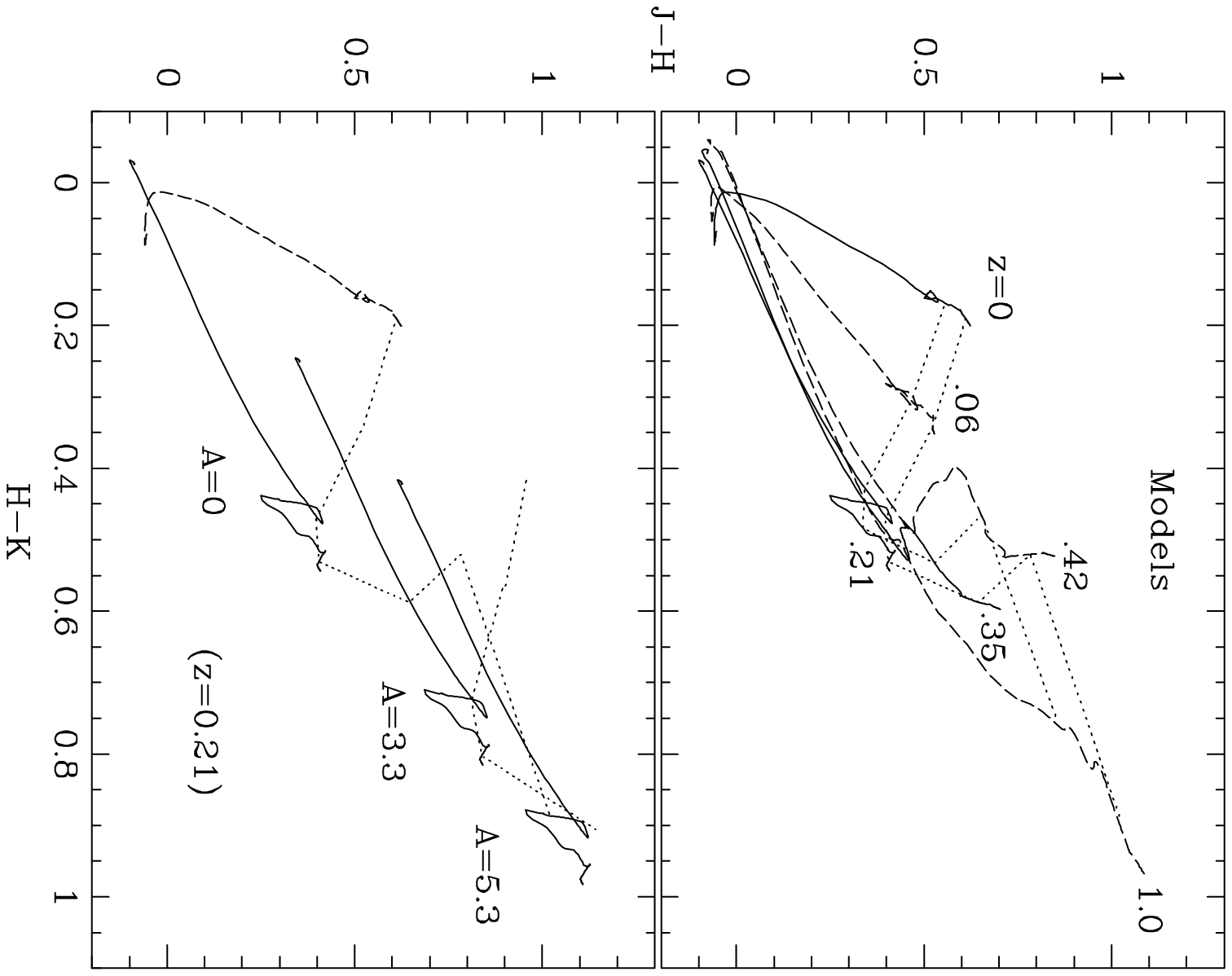]{Illustrative Gissel model tracks.
All tracks evolve from their
lower left to upper right, from age 0 to 20 Gyr. The initial population
is a 1 Gyr starburst, followed by passive evolution. Tracks are shown for
various redshifts covered by this program, in the upper panel. The dotted
lines show the loci of 2 and 10 Gyr age populations as redshift increases.
The lower panel shows the effects of reddening, with two values at z=0.21.
The dotted lines show the loci of 10 Gyr old populations with zero and
A$_V$=3.3 reddening. See text for more detail. \label{mod}}

\figcaption[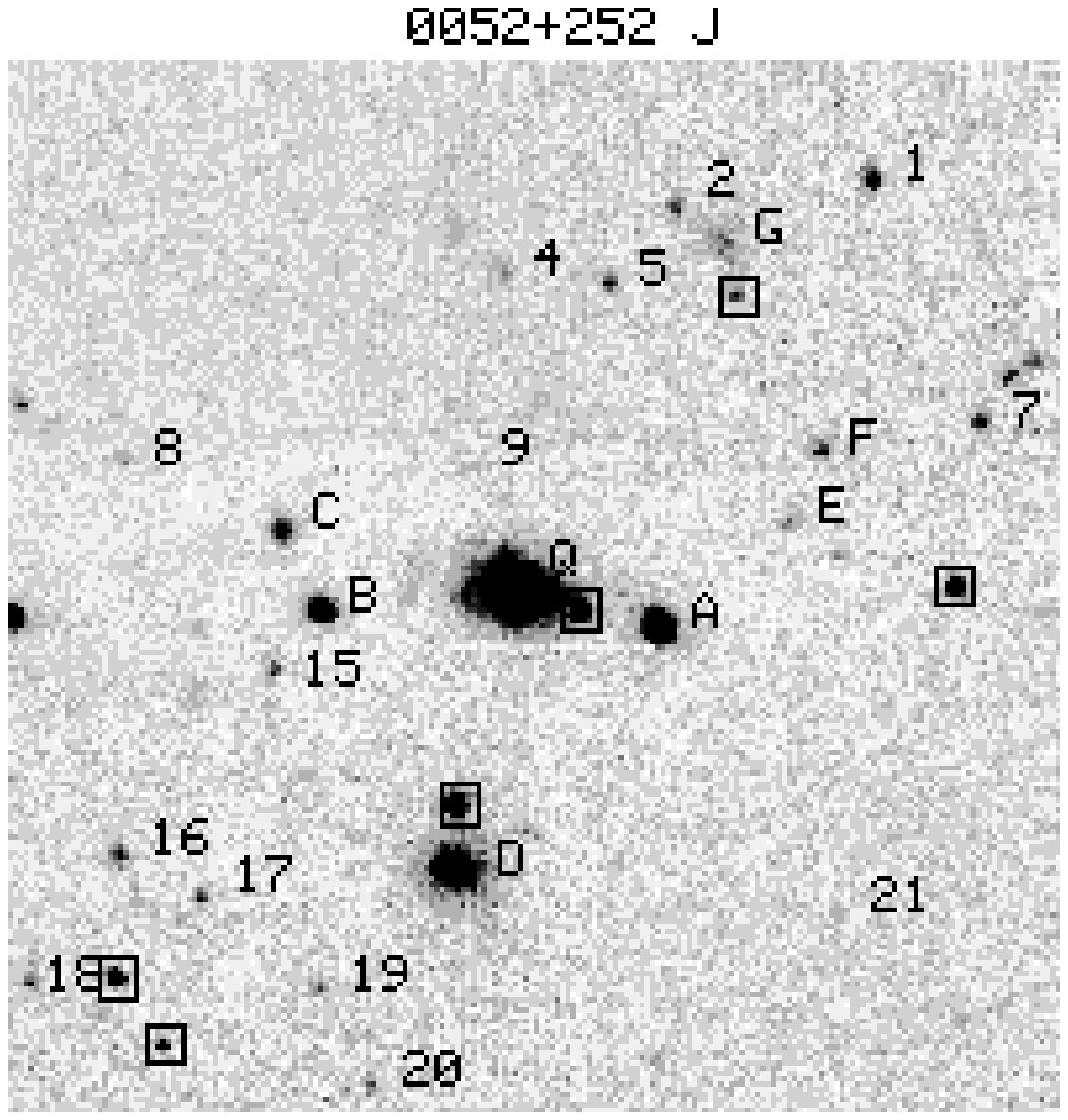]{The field of 0052+252 showing the location
of non-stellar objects measured in 2 colours with good precision (see text).
Stars are marked by squares. The sample is representative but not complete
as objects which are badly blended or which have poor signal or image
flaws in one colour are not included. Faint objects and stars are also 
excluded. \label{f00im}}

\figcaption[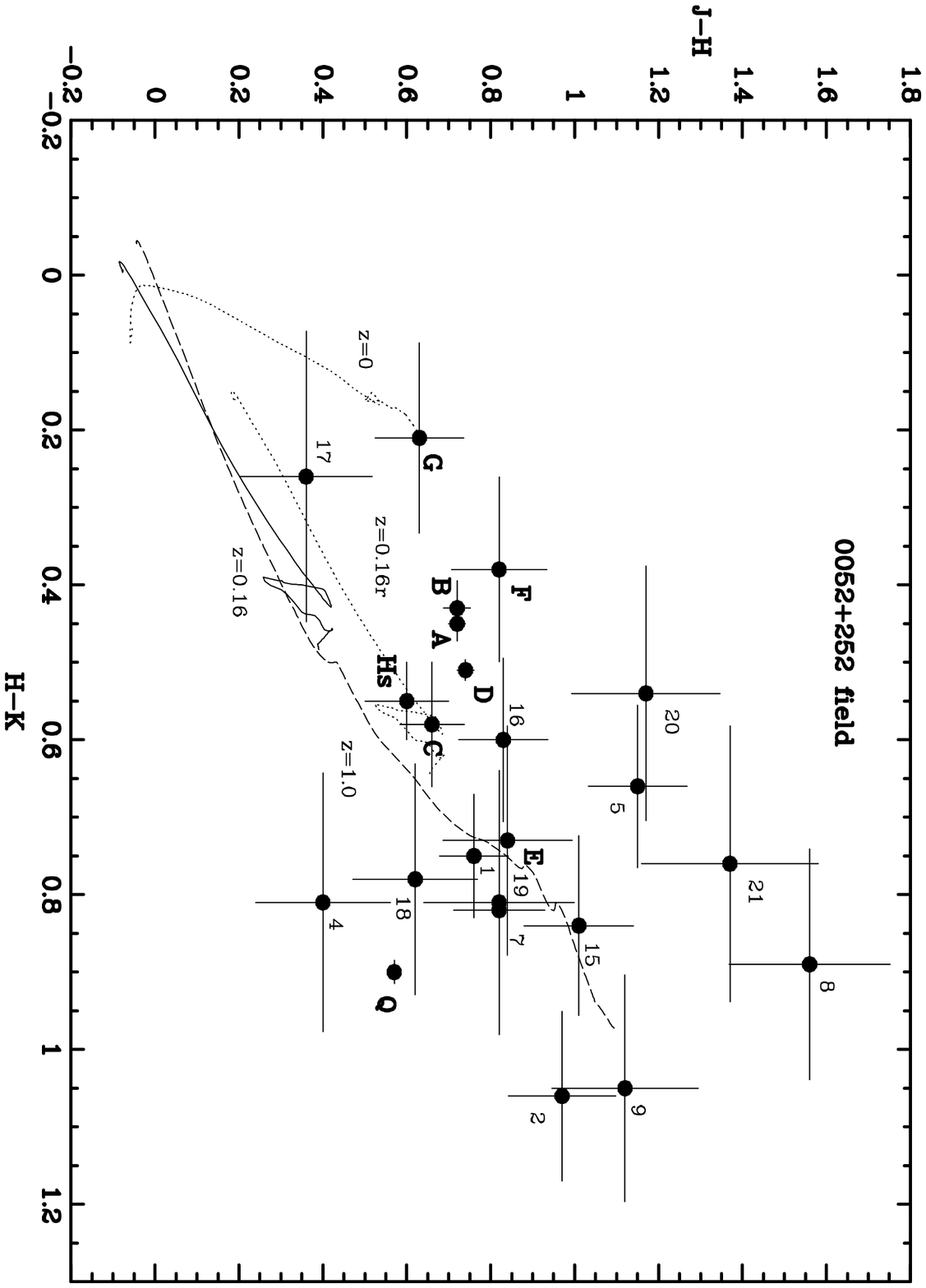]{The 2-colour diagram for galaxies in the
field of 0052+252. Error bars are 1$\sigma$ as explained in the text.
The QSO is marked Q and Hs indicates the host galaxy free of
nuclear contamination. Stellar population models are shown for the QSO 
redshift, unreddened and with A$_V$=2.1. We also show unreddened models for 
z=0 and z=1 for reference. Note the grouping of the bright galaxies and 
the host galaxy, suggesting some reddening or a systematic offset 
with respect to the models. \label{f00pl}}

\figcaption[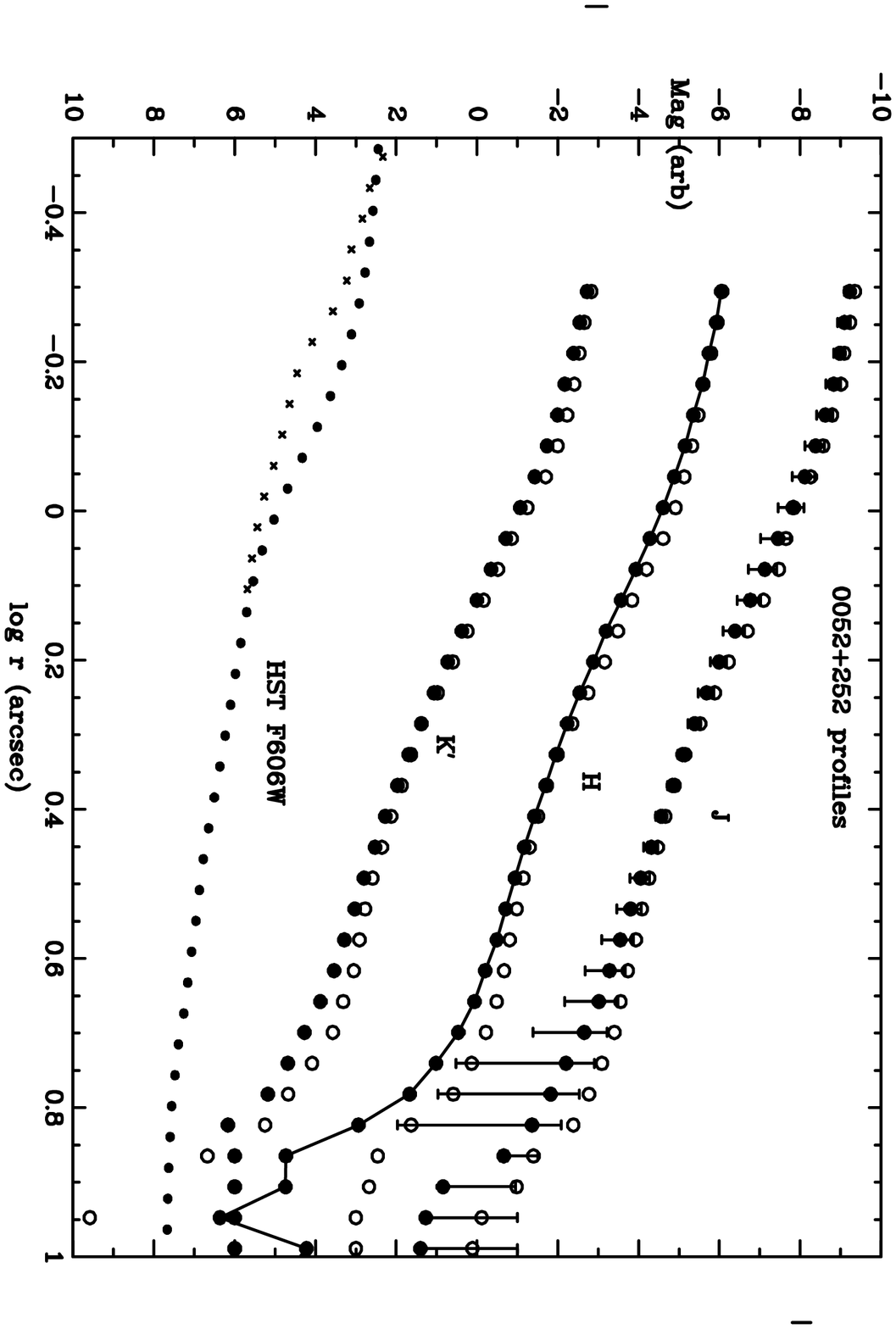]{Luminosity profiles of 0052+252 without nearby stars
and companions. Filled symbols are azimuthally averaged and open circles
are semi-major axis. One set of (typical) error bars are shown: missing
low bars at high r are large due to log y-scale. The HST image profile 
shows the difference at shorted wavelengths: the crosses are the raw image
and the dots are resampled to match our NIR data. \label{f00pr}}

\figcaption[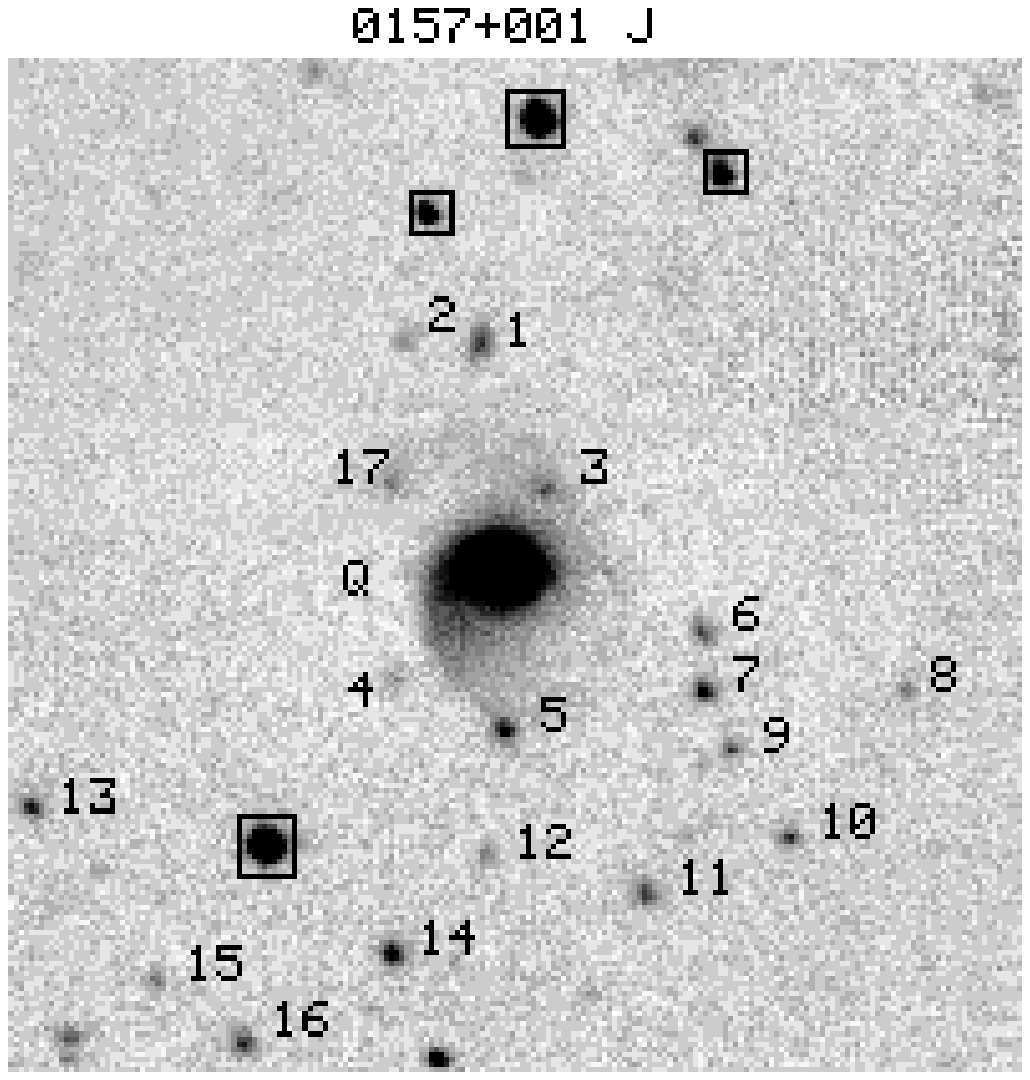]{J-band image of 0157+001 showing stars and measured
galaxies, as in Figure~\ref{f00im}. Features 3, 5, 17 all lie within the
host galaxy spiral arms. \label{f01im}}

\figcaption[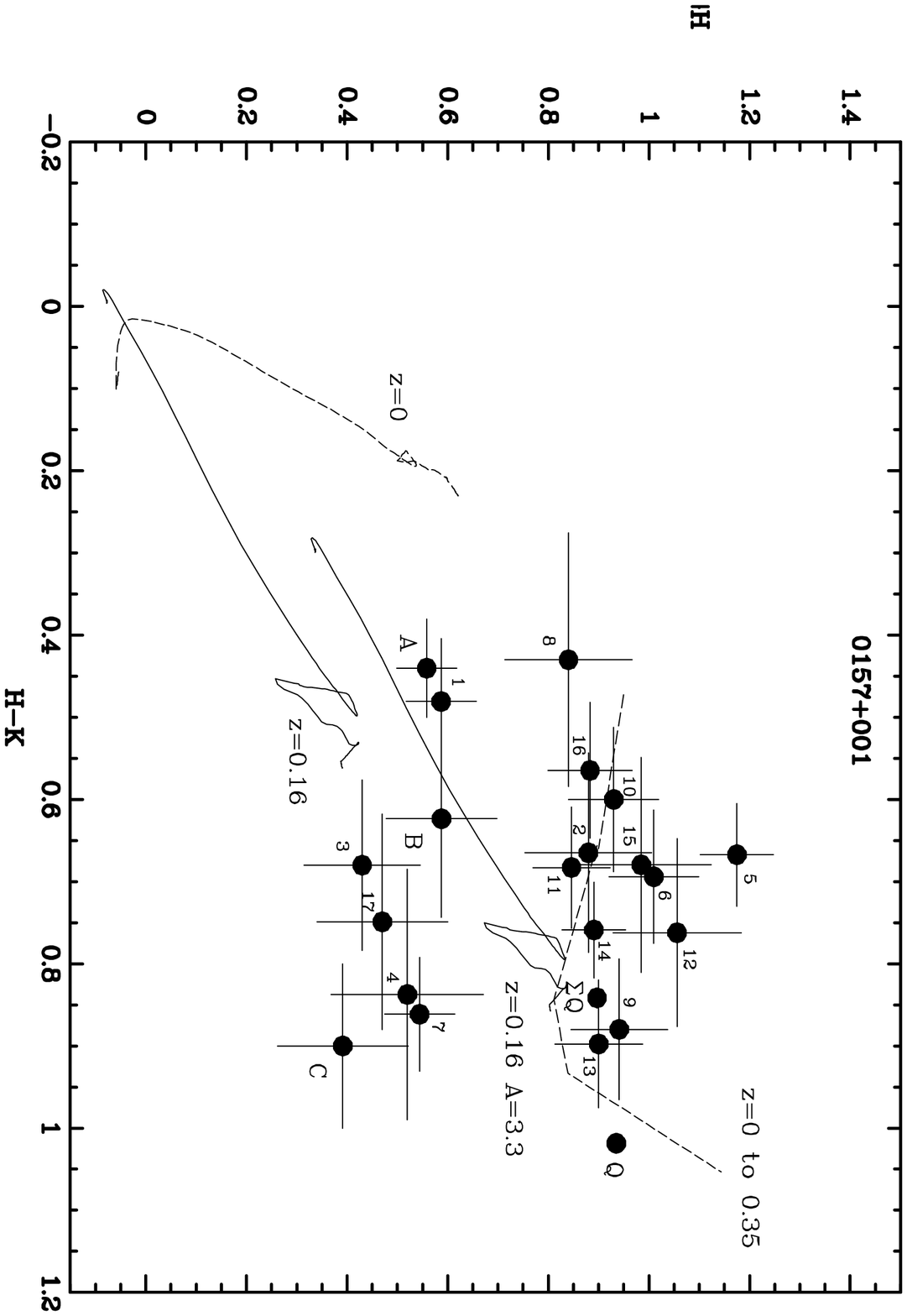]{The 2-colour diagram of 0157+001, as 
Figure~\ref{f00pl}. Q marks the QSO nucleus and $\Sigma$Q includes the host
galaxy. A, B, C are regions within the host galaxy, as described in the text.
Models are shown as in Figure~\ref{f00pl}, and the locus of age 10Gyr
reddened populations for z ranging from 0 to 0.35. The host galaxy regions
sampled and most of the nearest measured objects appear to have young
populations. \label{f01pl}}

\figcaption[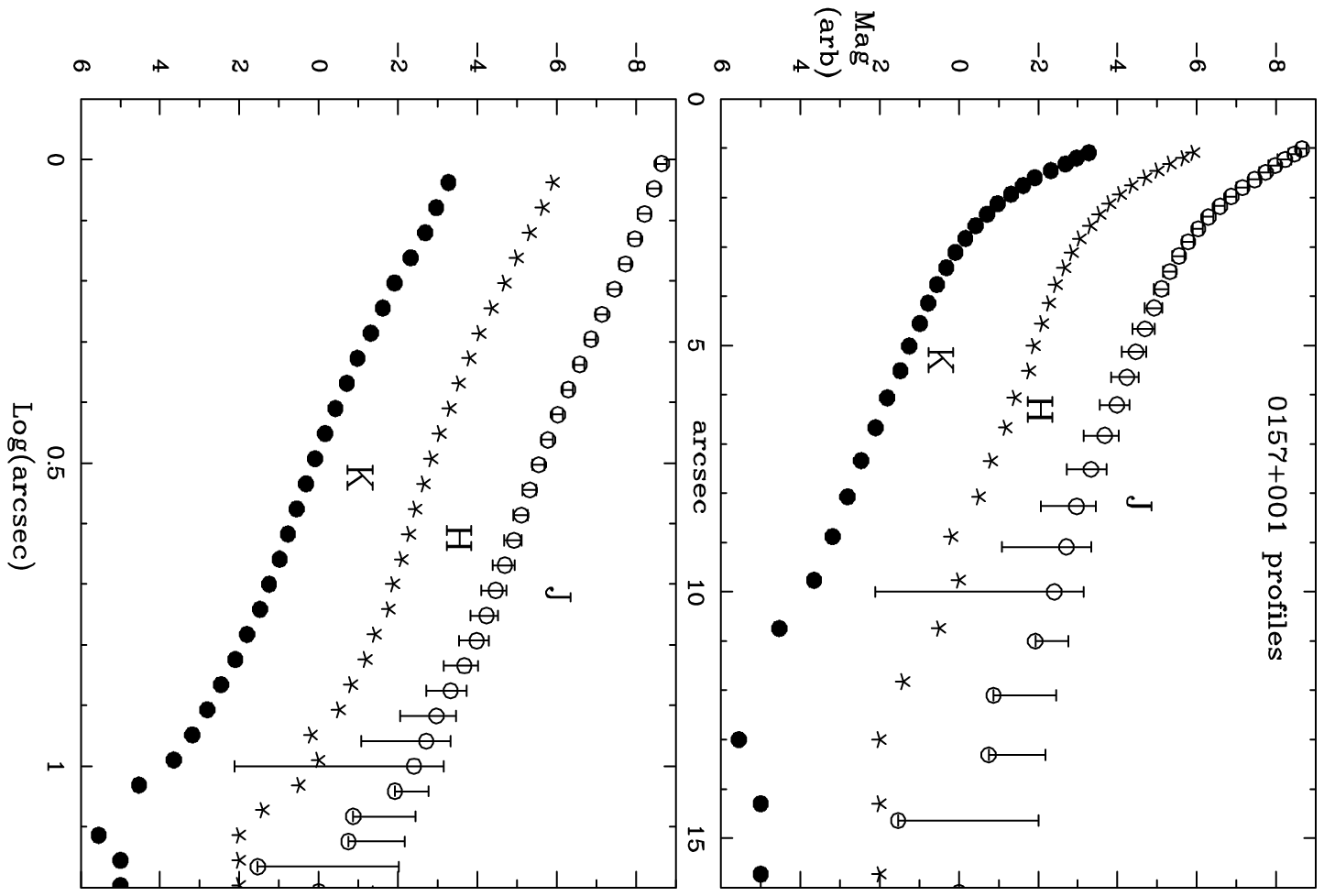]{Azimuthally averaged luminosity profiles for 0157+001.
Error bars shown for J-band are typical of H, K. The profile is composite,
as discussed in the text. \label{f01pr}}

\figcaption[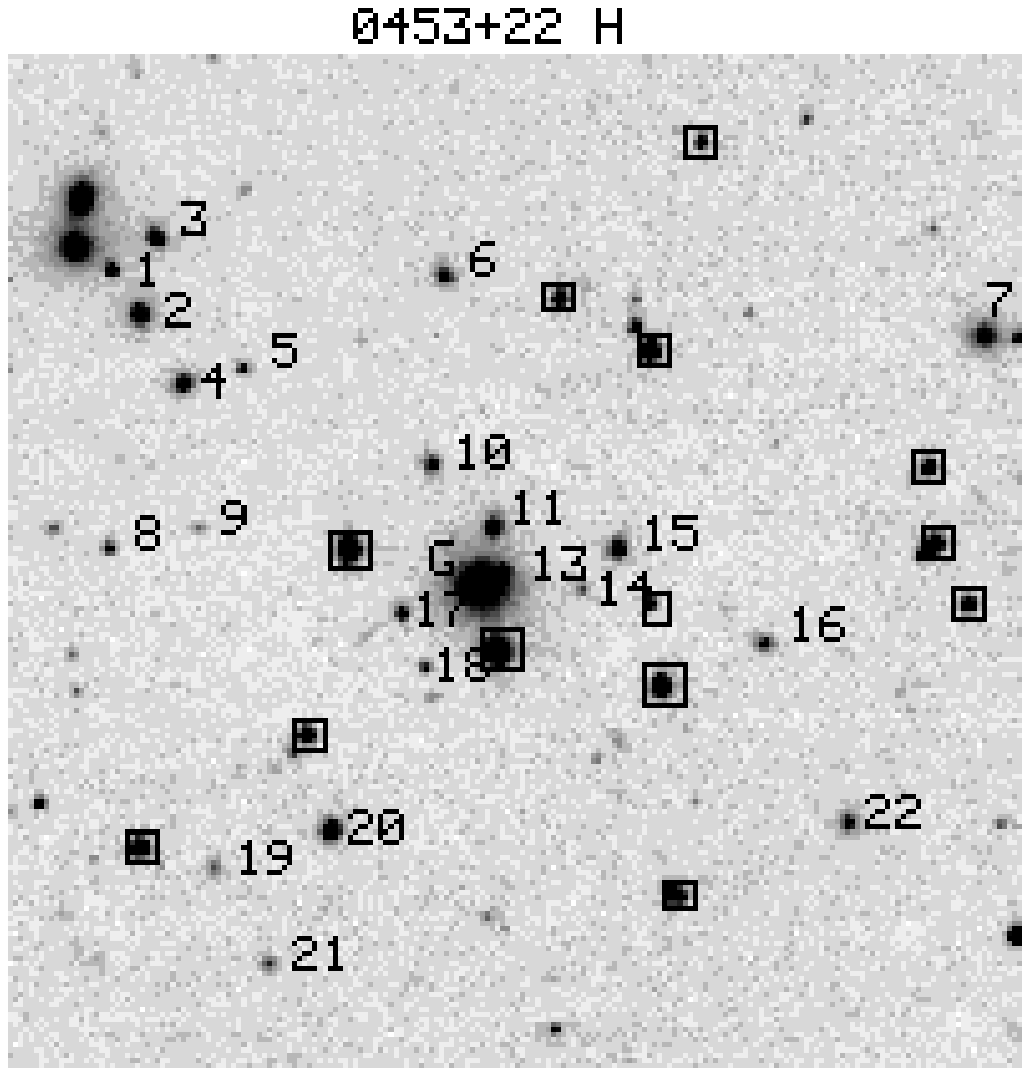]{H-band image of 0453+22=3C132 field with stars 
and measured galaxies. \label{f04im}}

\figcaption[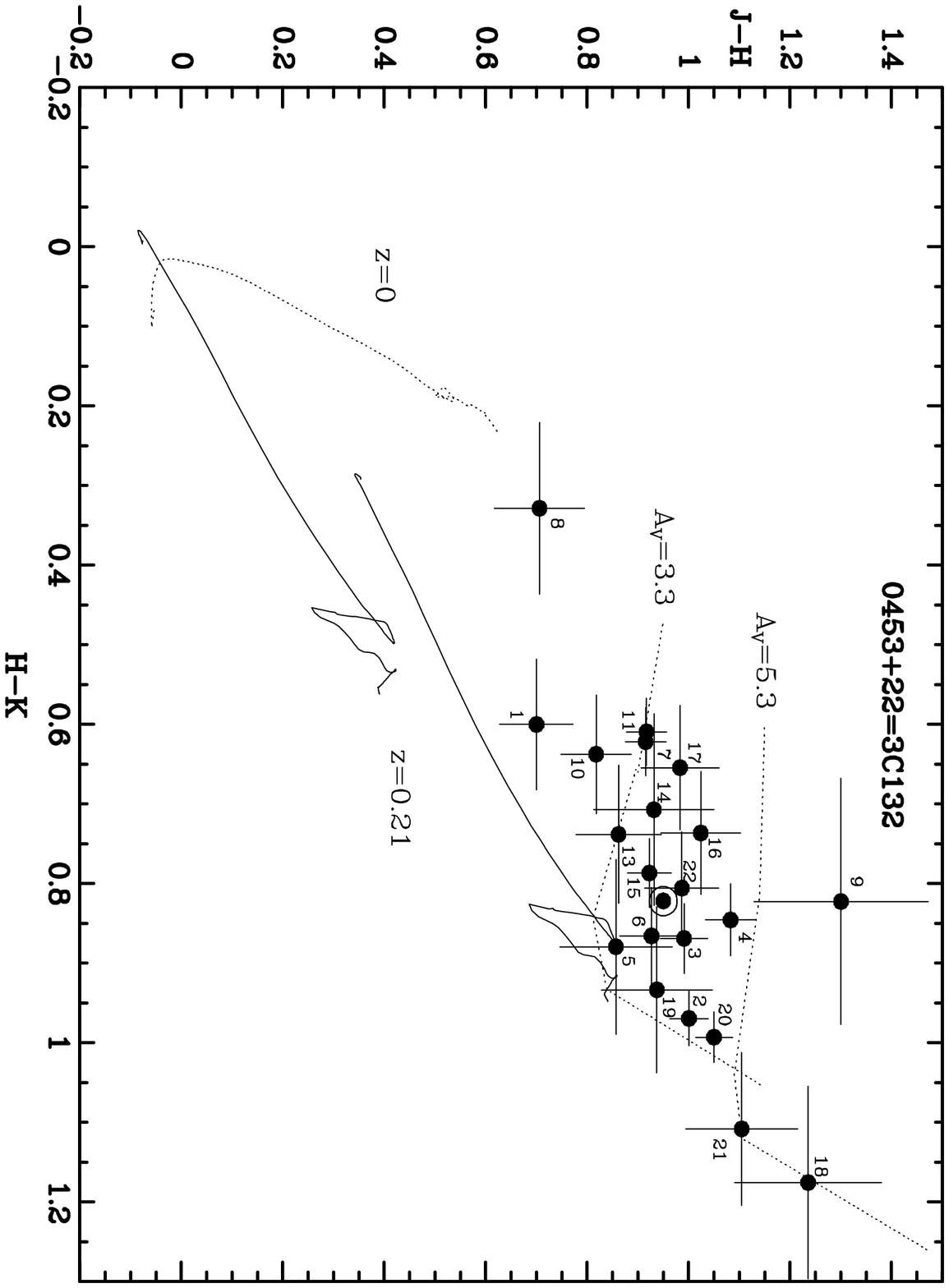]{2-colour plot of objects in 0453+22 field. The circled
point is the radio galaxy. The models for the galaxy redshift are for zero
and A$_V$=3.3 reddening. Loci for a 10Gyr age population with redshift are
shown for two values of reddening. Note the strong clustering of objects
near the radio galaxy. \label{f04pl}}

\figcaption[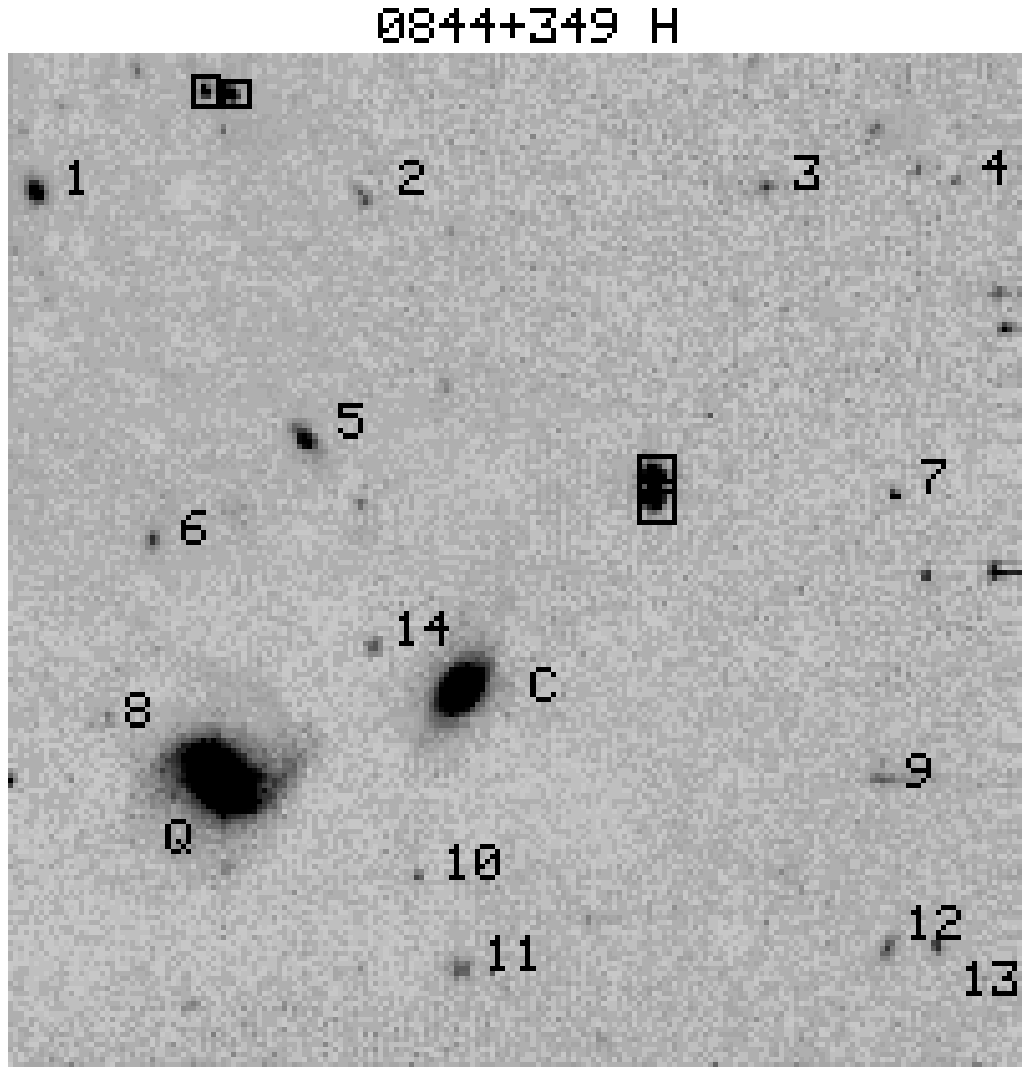]{H-band image of 0844+349 field, with stars and measured
objects. The QSO is marked Q and its interacting companion is C.
\label{f08im}}

\figcaption[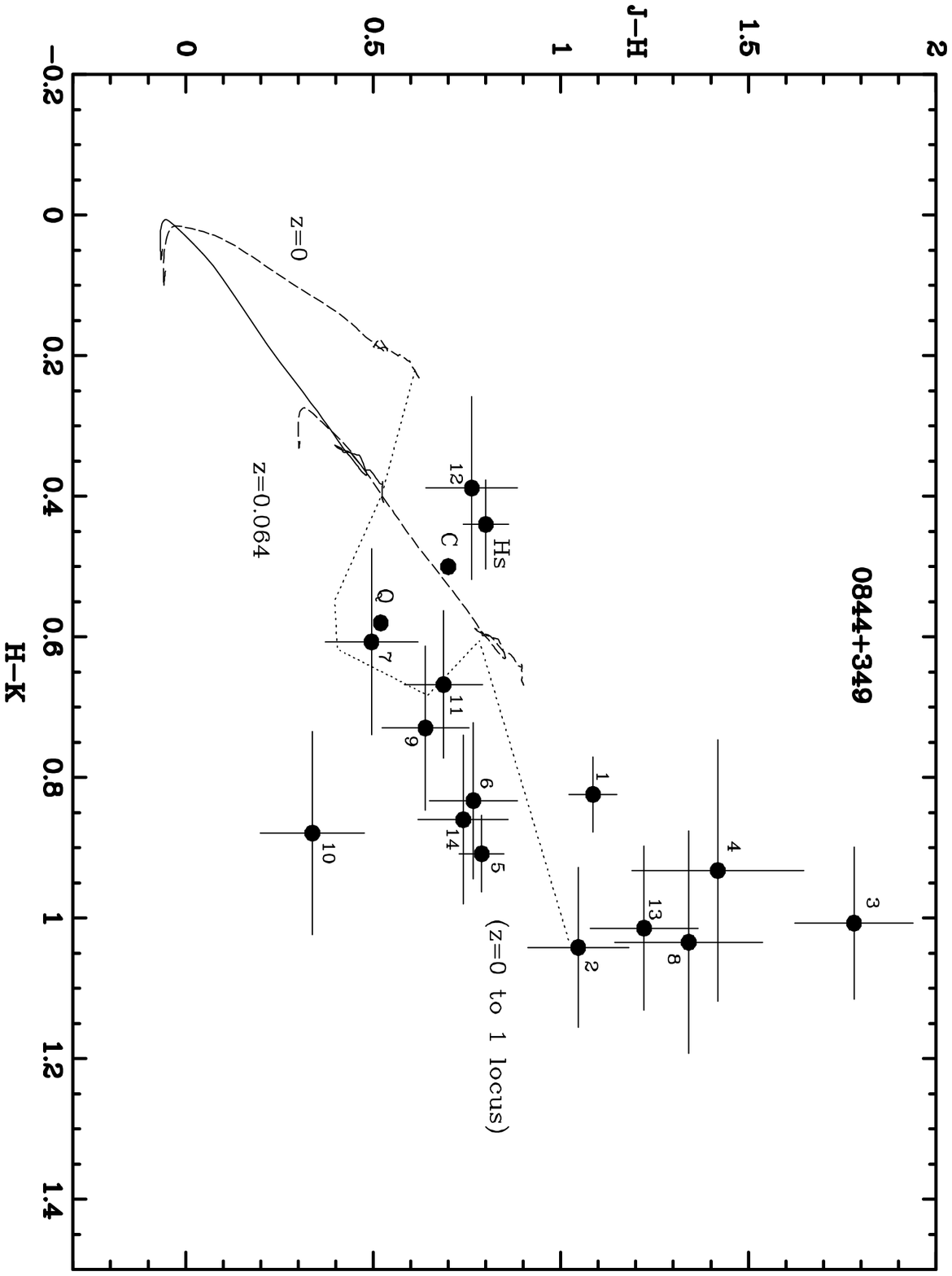]{2-colour plot of objects in 0844+349 field. Q refers
to the whole QSO plus host, while Hs shows the host galaxy outside the nucleus.
Note the similarity of colours of the host and the companion C. Otherwise,
there does not seem to be a significant group of faint galaxies around the 
QSO. Models are for the QSO 
redshift, and z=0, and the locus of unreddened 10 Gyr age populations is
shown for z=0 to 1.0. \label{f08pl}}

\figcaption[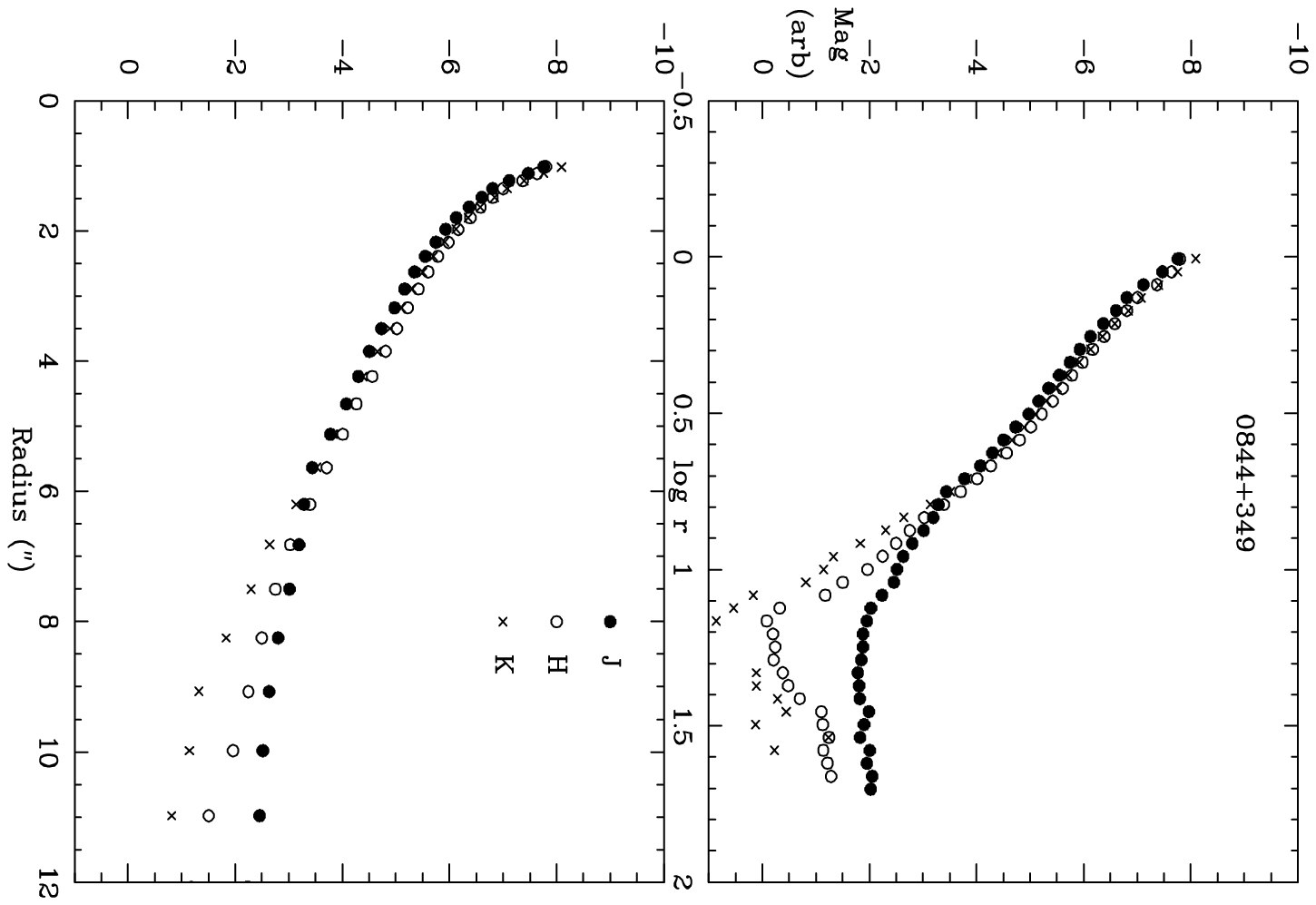]{Azimuthally averaged luminosity profiles of 0844+349,
showing its composite nature. Values at large radii are uncertain because
of sky frame noise and the bright nearby companion. \label{f08pr}}

\figcaption[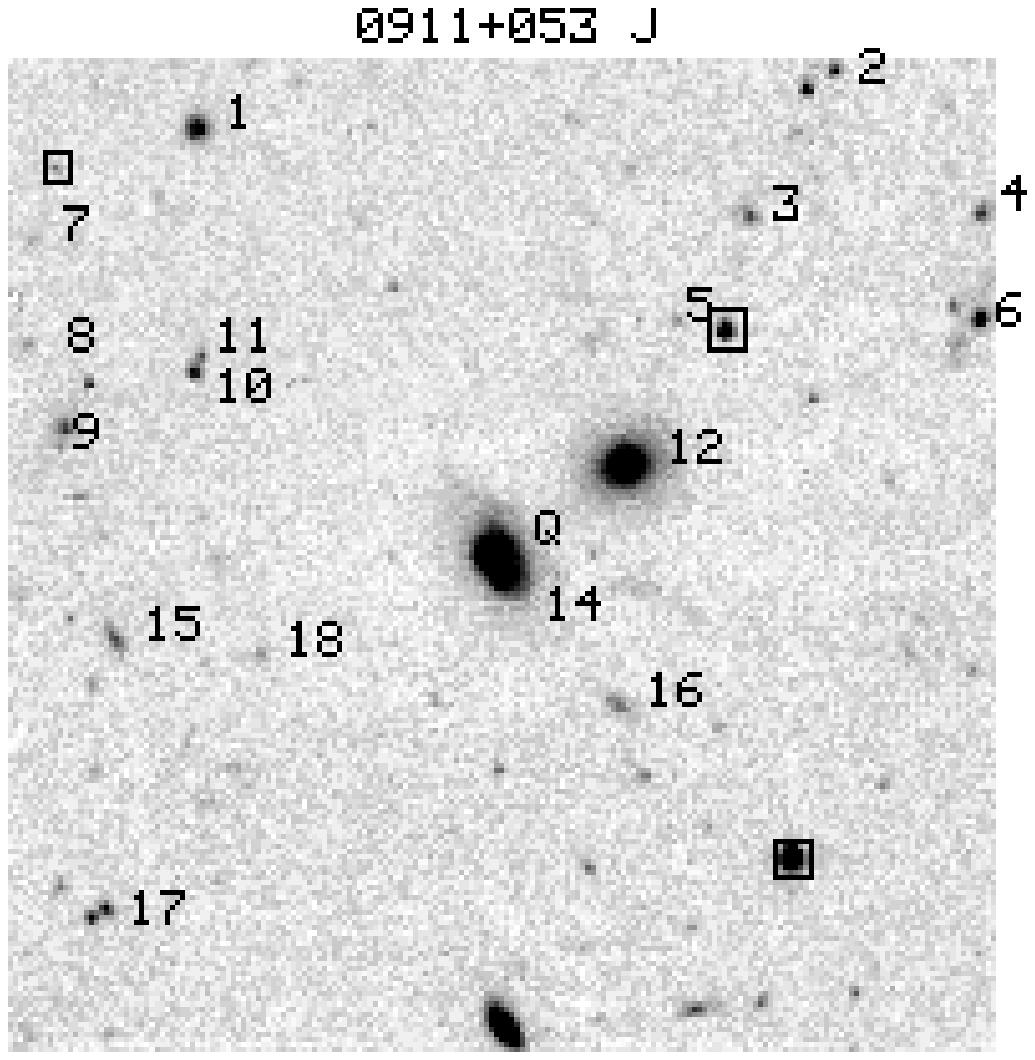]{J-band image of 0911+053 showing measured galaxies
and stars. Object 14 is a clearly separated second `nucleus' at $\sim$10\%
of the main peak flux. \label{f09im}}

\figcaption[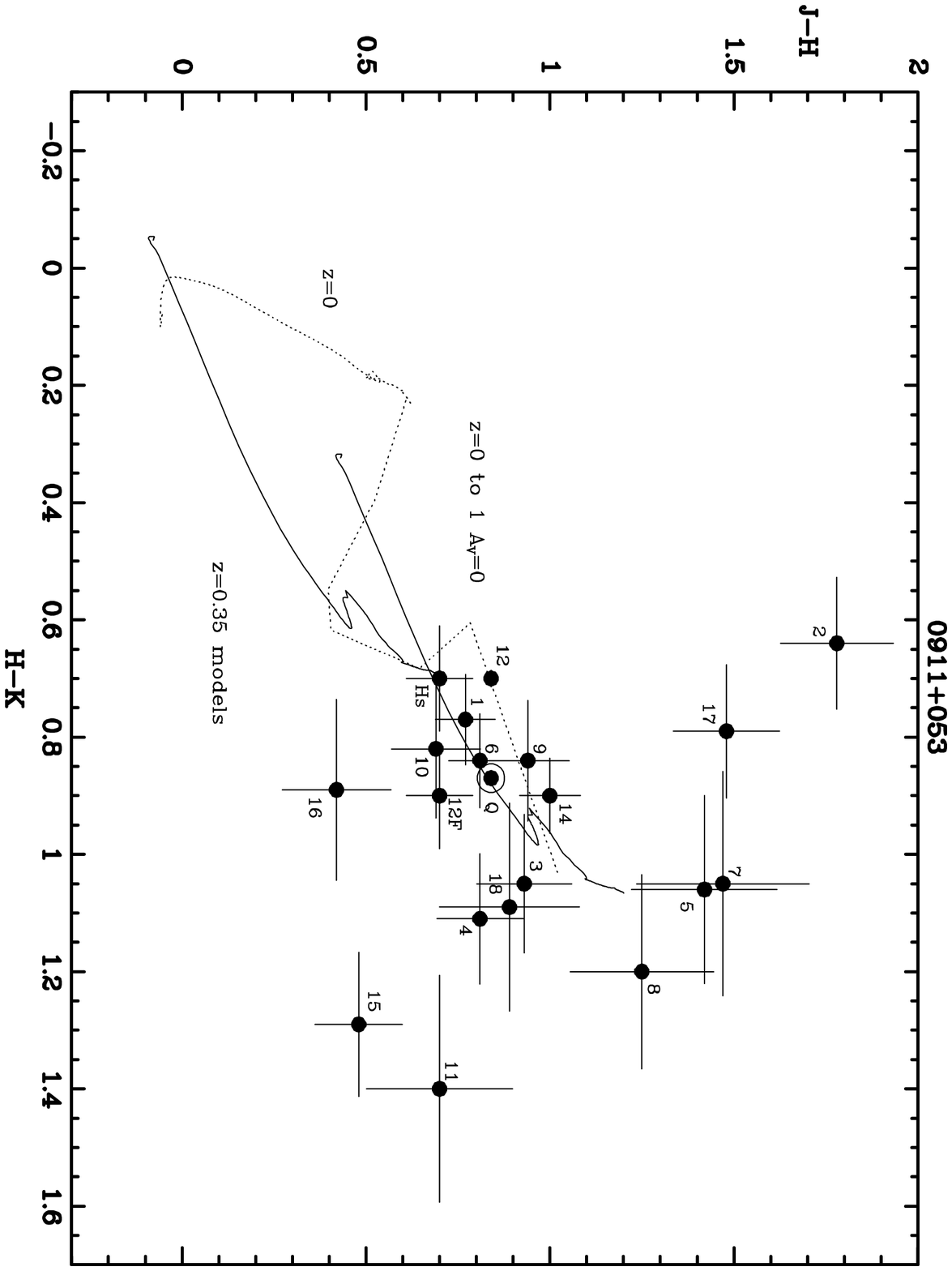]{2-colour plot of galaxies in 0911+053 field. The circled
point is the whole QSO, and Hs is the host galaxy away from the nucleus.
Note the group around the QSO position. The host galaxy may have a significant
younger population. Models at redshift near that of the QSO (0.30) are shown
for zero and A$_V$=3.3, and the locus of 10 Gyr age unreddened populations
for z=0 to 1.0. \label{f09pl}} 

\end{document}